\documentclass[aps,prb,superscriptaddress,twocolumn,showpacs]{revtex4}

\usepackage{graphicx}
\usepackage{amsmath}

\usepackage{epsfig}
\usepackage{color}
\usepackage{subfigure}

\begin{document}

\title{Andreev drag effect via magnetic quasiparticle focusing in 
normal-superconductor nanojunctions}

\author{P. K.~Polin\'ak}
\affiliation{Department of Physics, Lancaster University, Lancaster,
LA1 4YB, UK}
\author{C.~J.~Lambert}
\affiliation{Department of Physics, Lancaster University, Lancaster,
LA1 4YB, UK}
\author{J.~Koltai}
\affiliation{Department of Biological Physics, E{\"o}tv{\"o}s
University, H-1117 Budapest, P\'azm\'any P{\'e}ter s{\'e}t\'any 1/A, Hungary}
\author{J.~Cserti}
\affiliation{Department of Physics of Complex Systems, E{\"o}tv{\"o}s
University, H-1117 Budapest, P\'azm\'any P{\'e}ter s{\'e}t\'any 1/A, Hungary}

\begin{abstract}
We study a new hybrid normal-superconductor (NS) $\pi$-junction 
in which the non-local current can be orders of magnitude larger than
that in earlier proposed systems. 
We calculate the electronic transport of this NS hybrid 
when an external magnetic field is applied. 
It is shown that the non-local current exhibits oscillations as a
function of the magnetic field, making the effect tunable with the field. 
The underlying classical dynamics is qualitatively discussed. 
\end{abstract}

\pacs{74.45.+c, 75.47.Jn}

\maketitle

Electron-transport properties of normal-superconductor hybrid
nanostructures have been the subject of extensive theoretical and experimental~\cite{experiments} attention. 
Experiments carried out on nanostrucures containing ferromagnets (F) and superconductors (S) 
reveal novel features, not present in normal-metal/superconductor (N/S) 
junctions, due to the suppression of electron-hole correlations 
in the ferromagnet. 
When spin-flip processes are
absent, further effects are predicted, including the suppression
of conventional giant magnetoresistance in diffusive
magnetic multilayers~\cite{gmr} and the appearance of 
{\em non-local currents} when two fully-polarized ferromagnetic 
wires with opposite polarizations make contact with a spin-singlet
superconductor~\cite{feinberg}. 
The latter effect, also called {\em the Andreev drag effect},  
has been highlighted, because of interest 
in the possibility of generating entangled pairs of electrons 
at an N-S interface~\cite{tangle1}. 
A recent study of such a junction in the tunneling 
limit~\cite{feinberg} predicts that the magnitude of
the non-local current decreases exponentially as $\exp{(-2L/\pi\xi_c)}$, 
where $\xi_c$ is the superconducting coherence length, and $L$ is the 
distance between the F contacts. 
The effect can be enhanced by inserting a diffusive normal 
conductor between the superconductor and the ferromagnetic contacts leads 
as shown in Ref.~\cite{sanchez,janos}.
The off-diagonal conductance~\cite{feinberg}, which is always negative 
in the normal case, can have a positive value  
of order the contact conductances of these systems.
However, the value of the off-diagonal conductance is determined 
by fixed material parameters, such as the polarization of the 
ferromagnets, and the spin-flip time in the normal diffusive conductor.   
Therefore, it is of interest to study alternative methods for material-independent tuning of the non-local
current.

\begin{figure}[htb]
\includegraphics[scale=0.27]{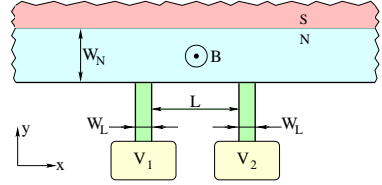}
\caption{\label{system}  
(color online) The hybrid N/S nanostucture consists of an infinitely long N/S 
ballistic waveguide, comprising a normal (N) metal of width $W_N$ coupled to a spin-singlet
superconductor (S) region of a width that is much larger than 
the superconducting coherence length $\xi_c$. 
To measure the non-local current, two ballistic normal leads, at
voltages $v_1$ and $v_2$, and with width $W_L$, separated 
by a distance $L$, are in contact with the normal waveguide. 
The left and right ends of the waveguide act as drains which absorb 
any quasi-particles exiting to the left or right. 
The magnetic field $B$ is applied perpendicular to the system
 (in our calculation $B>0$ corresponds to a field pointing out of the plane of the system).}
\end{figure}

In this work, we show that even in the absence of ferromagnetic contacts,
an enhanced Andreev drag effect is possible with the N/S structure 
shown in Fig.~\ref{system}. We shall demonstrate that the non-local current is enhanced by orders of
magnitude compared with the structure in which ferromagnetic leads
were used to detect the current~\cite{feinberg}. Moreover, the magnitude of the non-local current 
can be tuned by varying a magnetic field applied perpendicular to the system. 
The necessary field is much lower than 
the critical field of the superconductor. 

To calculate the non-local current we employ the current-voltage relation developed 
for normal/superconducting hybrid structures 
in the linear response limit~\cite{lam1}.
Assuming that the voltages $v_2$ at lead $2$, 
is the same as the voltage $v$ of the condensate potential, 
for the arrangement shown in Fig.~\ref{system} one finds
that the currents in lead $1$ and $2$ are 
\begin{subequations}
\begin{eqnarray}
I_1  &=& \frac{2e^2}{h} \, \left(N-R_0+R_a\right)(v_1-v)
\label{i1}, \\
I_2  &=& \frac{2e^2}{h} \, \left(T_a-T_0\right) (v_1-v),  
\label{i2}  
\end{eqnarray}
\label{i-v:eq}%
\end{subequations}%
where $v_1$ is the voltage at lead $1$ and $N$ is the number of open 
scattering channels in the normal leads of width $W_L$.  
Here $R_0$ ($T_0$ ) are the reflection (transmission) coefficients  
for an electron from lead $1$ to be reflected (transmitted) to lead $1$
($2$), and  $R_a$ ($T_a$ ) are the  Andreev reflection (transmission) 
coefficients for an electron from lead $1$ to be reflected (transmitted) 
to lead $1$ ($2$)  as a hole. $R_a$ and $R_0$ satisfy the inequality 
$N-R_0+R_a \ge 0$, thus, $I_1$ is always positive for positive $v_1-v$.   
All coefficients are evaluated at the Fermi energy using an exact scattering matrix formalism.

It is easy to see from Eq.~(\ref{i-v:eq}) that whenever 
Andreev transmission dominates normal transmission (ie $T_a > T_0$) 
the currents $I_1$ and $I_2$ have the same signs, ie a current in lead $1$ 
induces a current in lead $2$ flowing in the same direction. In semi-classical 
point of view, this means that hole like quasi-particles leave the system at lead $2$.
This is the Andreev drag effect.
On the other hand,  in the case when the normal transmission is larger than the Andreev
transmission (ie $T_a < T_0$), electron-like quasi-particles 
leave the system through lead $2$, yielding a current flowing opposite to
the direction of the current in lead $1$. 
  
In what follows now, we show that for the system depicted in Fig.~\ref{system} 
the ratio  $T_a/T_0$ can be tuned by an applied magnetic field.
To this end, we calculate the transmission coefficients for the
system using the Green's function
technique~\cite{sanvito} developed for discrete lattice.  
Each site is labelled by discrete lattice coordinates
$(x,y)$ and possesses particle (hole) degrees of freedom
$\psi^{e(h)}(x,y)$. 
The magnetic field is incorporated via a Peierls
substitution.
In the presence of local s-wave pairing described by a
superconducting order parameter $\Delta(x,y)$, 
the Bogoliubov-de Gennes equation~\cite{BdG-eq} (BdG) 
for the retarded Green's function takes the form
\begin{subequations}
\begin{equation}
\left(
\begin{array}{cc}
\mathbf{H}-E & \Delta \\
\Delta^{\star} & - \mathbf{H}^{\star}+E
\end{array}
\right)
\left(
\begin{array}{cc}
G^{ee}  & G^{eh}  \\
G^{he}  & G^{hh} 
\end{array}
\right)=-\left(
\begin{array}{cc}
\mathbf{1} & \mathbf{0}\\
\mathbf{0} & \mathbf{1}
\end{array}
\right),\label{BdG}
\end{equation}
where the components of $\bf{H}$ are 
\begin{eqnarray}
\mathbf{H}_{x,x^\prime,y,y^\prime} & = & 
\left[\epsilon_0 - E_F\right]\delta_{x,x^\prime}\delta_{y,y^\prime} - 
\sum_{n_x} \gamma_x \delta_{x+n_x,x^\prime}\delta_{y,y^\prime} - \nonumber \\
& - & \sum_{n_y} \gamma_y \delta_{x,x^\prime}\delta_{y+n_y,y^\prime}.  
\end{eqnarray}
\end{subequations}
Here $E_F$ is the Fermi-energy, and $n_x$ and $n_y$ are the nearest neighbors of $(x,y)$ 
in the $x$ and $y$ direction, respectively.

Within the Landau-gauge with a vector potential in the $x$-direction,
$\gamma_x=\gamma_0 e^{i\theta(y)}$, $\gamma_y=\gamma_0$, where $\gamma_0$ 
is the hopping parameter without magnetic field.  
The phase $\theta(y)$ for the Peierls substitution  
is zero in the superconducting region, and it is given by 
$\theta(y)=Ba^2\pi(W_N-y)/\Phi_0$ in the normal region, 
where $a$ is the lattice constant~\cite{parameters:dat}, 
and $\Phi_0=h/2e$ is the flux quantum.  
This choice of gauge results in a uniform magnetic field $B$ 
in the normal region, and zero magnetic field in the S region, 
while the translation invariance in the $x$ direction is preserved.
The order parameter is assumed to be a step function~\cite{McMillan,Plehn}, 
ie constant $\Delta_0$ in the S region and zero otherwise. 
The phase $\theta$ is set to $\theta_{\rm lead}=Ba^2\pi W_N/\Phi_0$ in
the leads $1$ and $2$ to ensure the continuity of the vector
potential. 
The parameters of the Hamiltonian $H$ are chosen to model an 
experimentally-realizable situation in the quasiclassical 
regime, ie $W_N \gg {\rm Fermi\,\,\, wavelength}$~\cite{parameters:dat}. 

\begin{figure}[htb]
\includegraphics[scale=0.23]{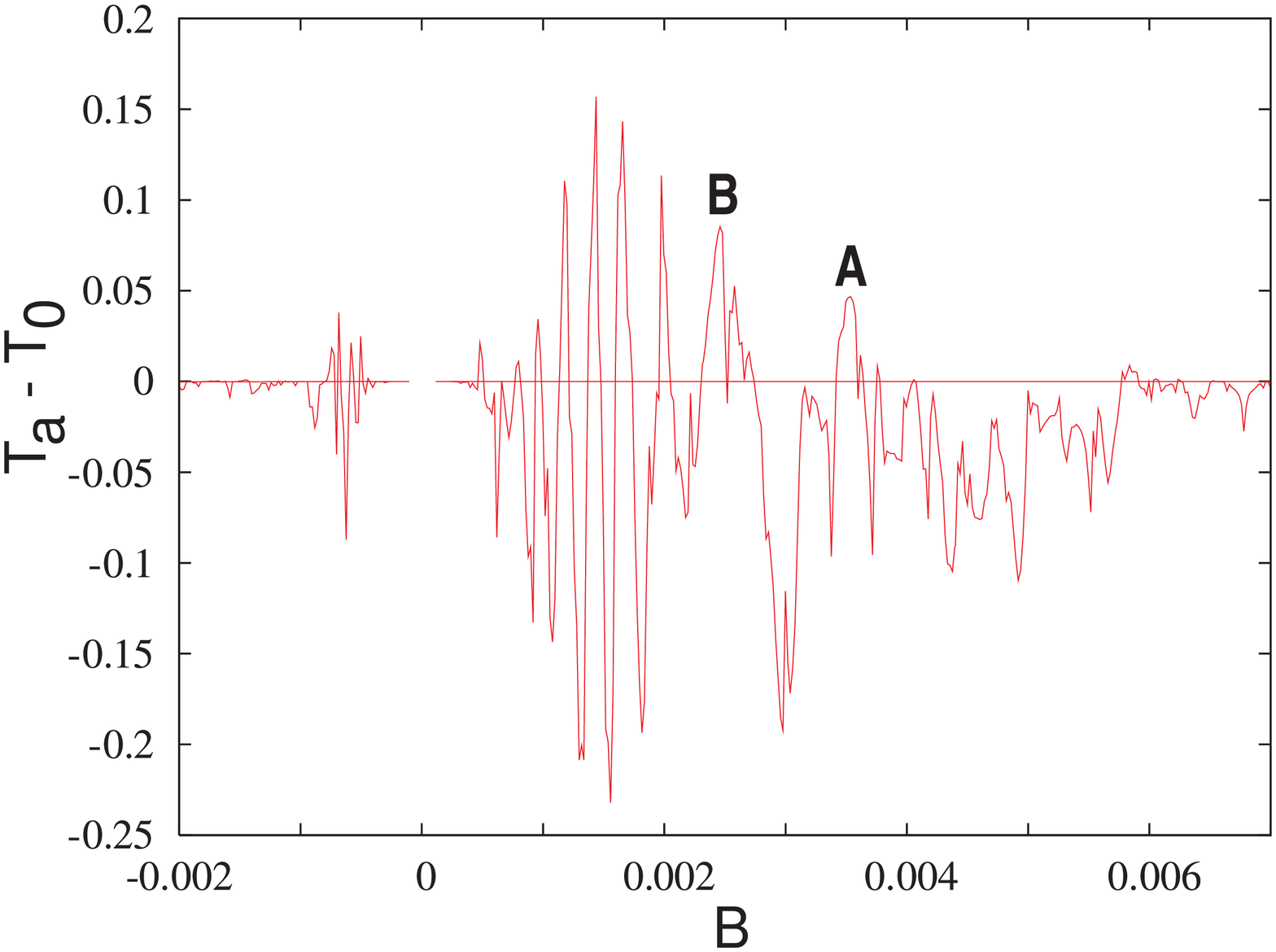}\vspace{-5mm}
\caption{\label{trans1} (color online) 
$T_a-T_0$ as a function of the magnetic field 
(in units of $\Phi_0/(2 a^2\pi)$) at the Fermi energy $E_F$. 
In lead $1$ only one mode was allowed. 
The wave functions at magnetic fields corresponding to 
letters $A$ and $B$ on the peaks of the curves will be shown in
Fig.~\ref{wfns}.} 
\end{figure}

From the Green's function and the scattering matrix for the
system, the transmission and reflection coefficients 
are calculated as a function of the magnetic field.
Our central result, shown in Fig.~\ref{trans1}, is that the difference between the Andreev and
normal transmission coefficients $T_a-T_0$ (which proportional to the measurable  current $I_2$
according to Eq.~(\ref{i2})) is an oscillating function of the magnetic field. 
Furthermore, since positive peaks correspond to pronounced Andreev drag effect  
and the heights of the
positive peaks are comparable with those of the negative peaks,
the non-local current can be as large 
as the normal current in our hybrid system. 

\begin{figure}[htb]
\includegraphics[scale=0.13]{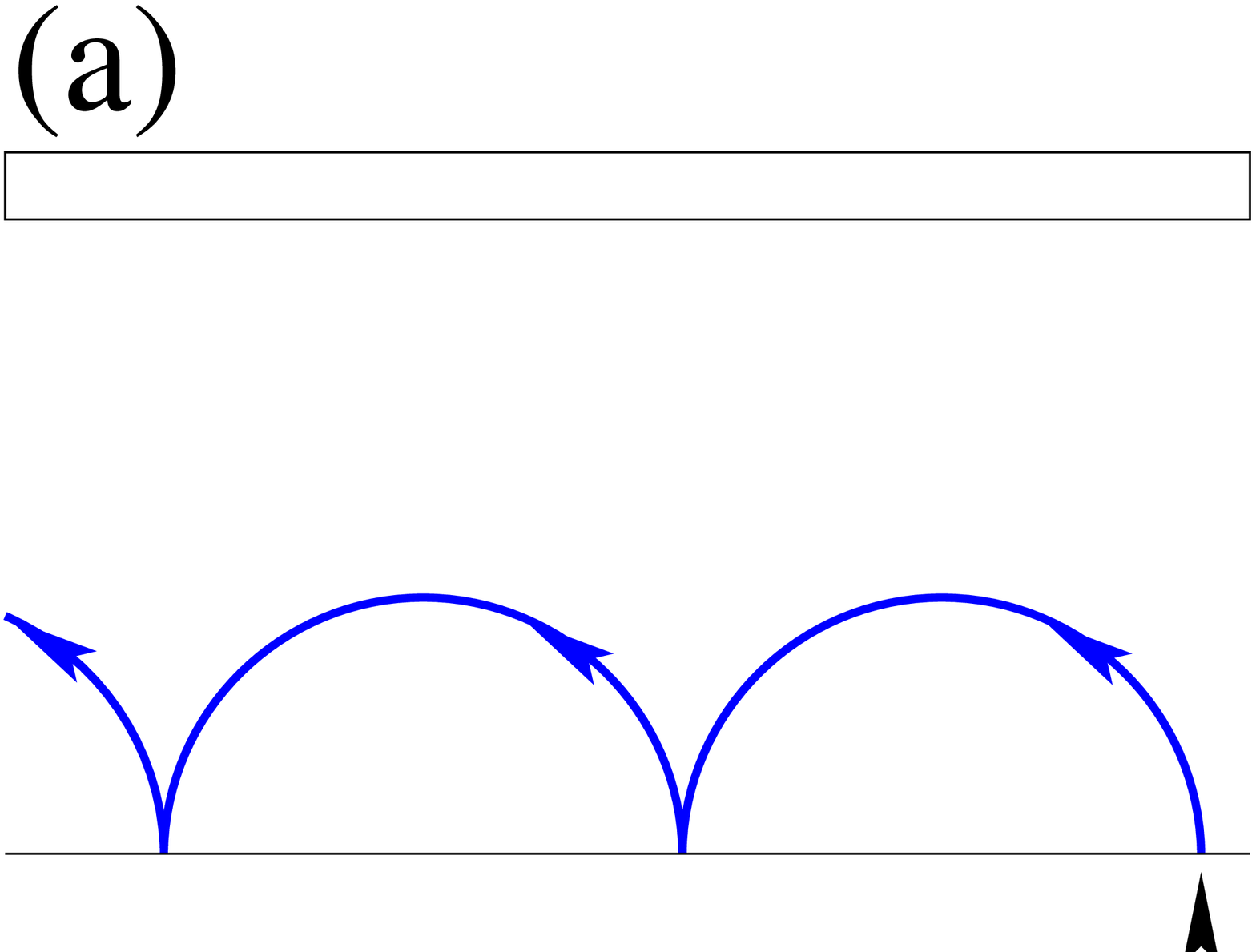}
\includegraphics[scale=0.13]{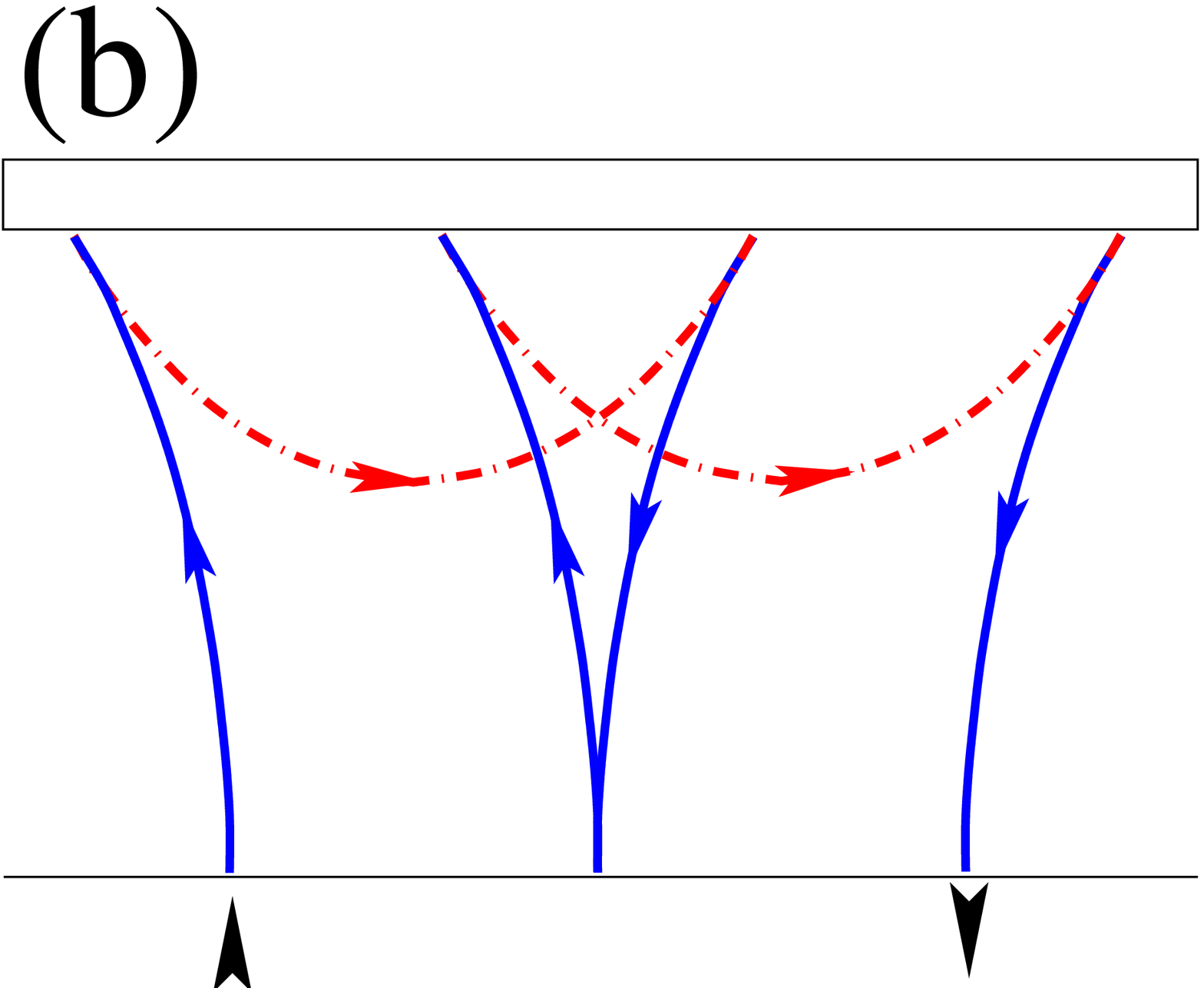}
\includegraphics[scale=0.13]{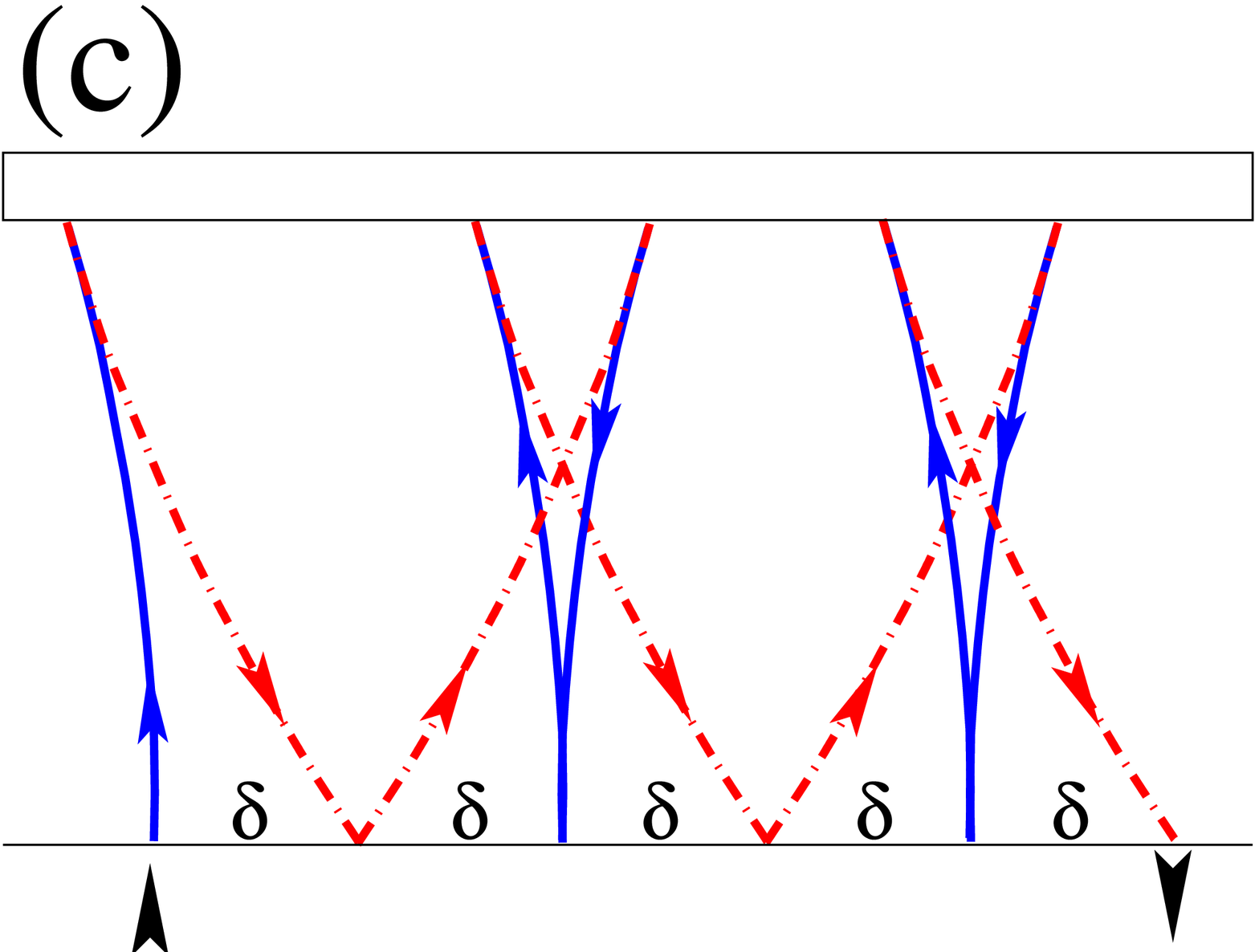}
\caption{\label{class-orbits:fig} (color online) 
Classical trajectories in a magnetic field. 
Figure (a) shows the case when the electron does not reach the N-S interface. 
New types of trajectories involving Andreev reflections are sketched on figures (b) and (c), 
for the cases when only the electron (b), both the electron
and the hole (c) can reach 
the side of the waveguide where the leads are attached. 
Blue (red) lines refer to the electron (hole). Electrons are injected perpendicularly 
into the waveguide at the positions marked by arrows pointing up. }
\end{figure}

A striking feature of Fig.~\ref{trans1} is that it is an asymmetric 
function of $B$. 
This can be understood qualitatively by tracing the classical cyclotron orbits of quasi-particles, bearing in mind that when electron-hole conversion occurs at the NS boundary, the 
chirality of the electron-like and the hole-like orbits is preserved and therefore
a geometrical construction for their classical
trajectories is different from that of normal systems~\cite{zulicke}. 
Examples of trajectories obtained from this new geometrical construction 
are plotted in Fig.~\ref{class-orbits:fig}.

For $B>0$ electrons injected from lead $1$ 
will follow classical orbits bending to the left. For a large enough 
$\left| B \right|$ these will exit to $x=-\infty$, without  
reaching the N-S interface, and impinging on lead $2$, as can be seen in Fig.~\ref{class-orbits:fig}a. 
Therefore for large positive $B$ all transmission coefficients from lead $1$ to lead $2$ 
vanish. Andreev reflection can occur if $\vert B\vert$ is sufficiently 
small to allow the electrons to reach the N-S interface. 
This condition is defined by $\vert B\vert <B_1$, where $B_1$ is the field for which the cyclotron radius $R_c = W_N$, 
where $R_c=\sqrt{2mE_F}/eB$. As shown in Figs.~\ref{class-orbits:fig}b-c, 
the transport direction is reversed compared to the normal case due to Andreev 
scattering, because even if the classical orbits are anti-clockwise, 
quasi-particle transport is to the right, resulting in quasi-particles impinging on lead 2. 
This is why the asymmetry in  Fig.~\ref{trans1} arises.
On the other hand, as shown in Fig.~\ref{class-orbits:fig}b, 
if $R_c$ is not sufficiently large, there is no drag effect, because the trajectories of the holes do 
not hit the side of the waveguide to which the leads are attached. 

\begin{figure}[h]
\includegraphics[scale=0.55]{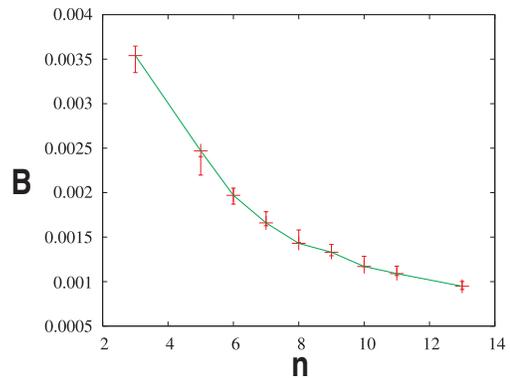}\vspace{-2mm}
\caption{\label{error:fig}  
The range of $B_n$ shown as a vertical bar for each $n$, together with those 
values of magnetic field at which we obtained peaks in $T_a$ from the quantum transport calculation. 
(Green line is connecting these peaks, but only for guiding the eyes). \vspace{-2mm}}
\end{figure}

Andreev drag effect only occurs  
for $\vert B \vert < B_{\rm max}$, where the maximum field 
$B_{\rm max}$ is determined from the condition $R_c \ge 2 W_N $.
By appropriate choice of the width $W_N$ of the normal part of the 
waveguide, $B_{\rm max}$ can be much less than the critical field of the 
superconductor. 
The trajectory relevant for this case is shown in Fig.~\ref{class-orbits:fig}c. 
On the normal side of the waveguide, normal quasi-particle reflections 
alternate between electrons and holes, separated by equal distances $\delta$.
Assuming that the electrons are injected perpendicularly into the waveguide, 
simple geometrical considerations give the following condition for  maxima in $T_a$:
\begin{equation}
L= (2n + 1)\delta,
\label{cond:eq}
\end{equation}
where $n$ is an integer counting the number of normal reflections of 
the hole at the side of the normal waveguide to which the leads
are attached, and  $\delta =  2 \sqrt{R_c^{2}-W_N^{2}} - \sqrt{R_c^{2}-4W_N^{2}} - R_c$. 
From Eq.~(\ref{cond:eq}) one can find a magnetic field $B_n$ for each $n$.
The peaks in $T_a$ can be expected at $B_n$.
Taking into account the finite widths of the two leads we calculated
the ranges of $B$ for each $n$ in which a peak in $T_a$ should be found, 
which corresponds to the range of $B$ for which a classical trajectory 
of the hole hits the finite-width interface of lead $2$. 
In Fig.~\ref{error:fig} we plotted the ranges of $B_n$ as vertical bars
together with those values of magnetic field at which we obtained
peaks from the exact quantum calculations.
One can see that the agreement between the quantum and the classical 
calculation is reasonable. 
Thus, this simple classical argument can be used to estimate 
the magnetic field needed to obtain enhanced Andreev drag.

\begin{figure}[htb]
\includegraphics[scale=0.41]{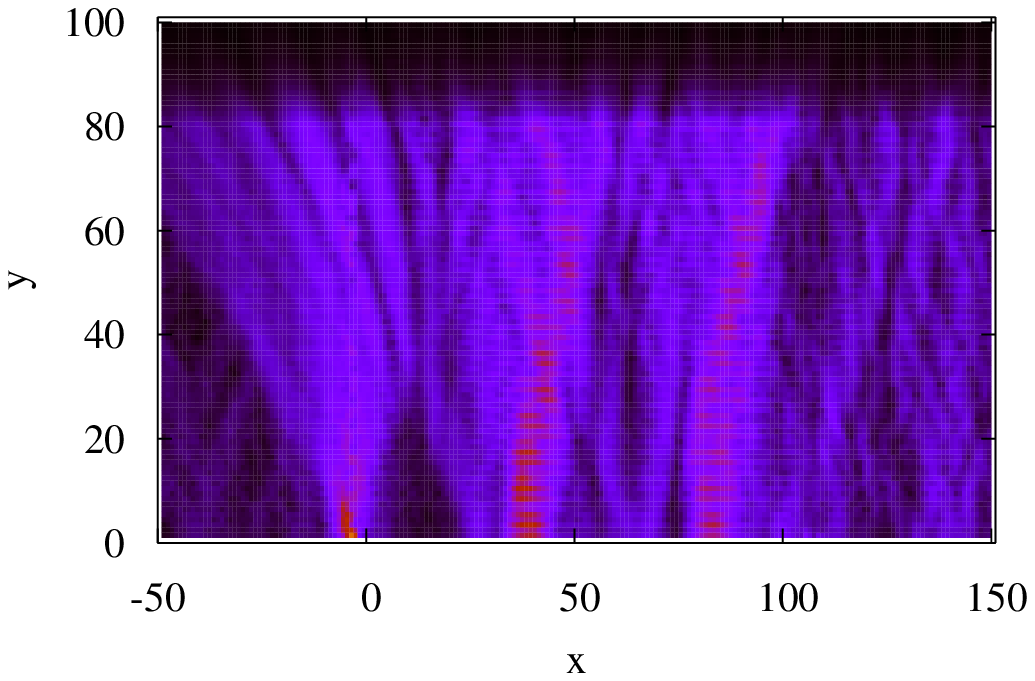}
\includegraphics[scale=0.41]{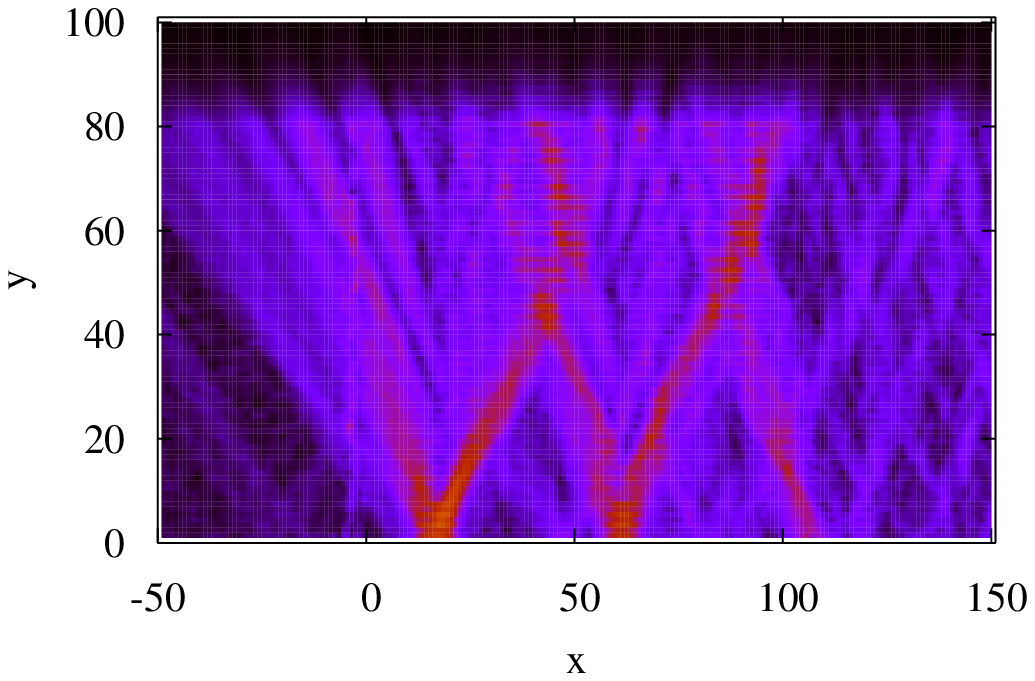}\vspace{2mm}
\includegraphics[scale=0.41]{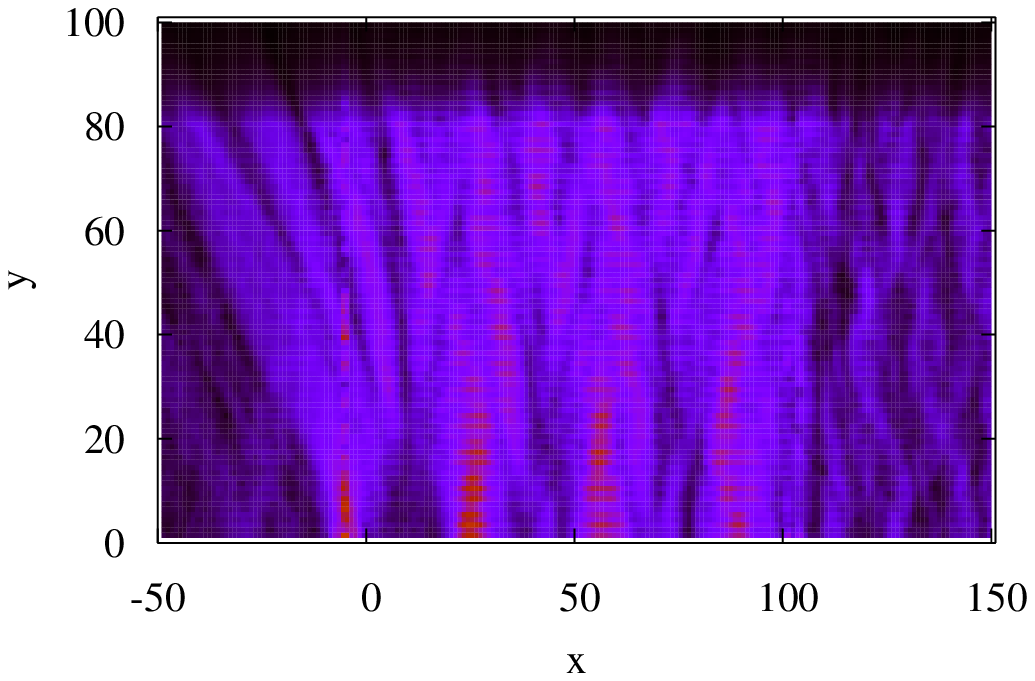}
\includegraphics[scale=0.41]{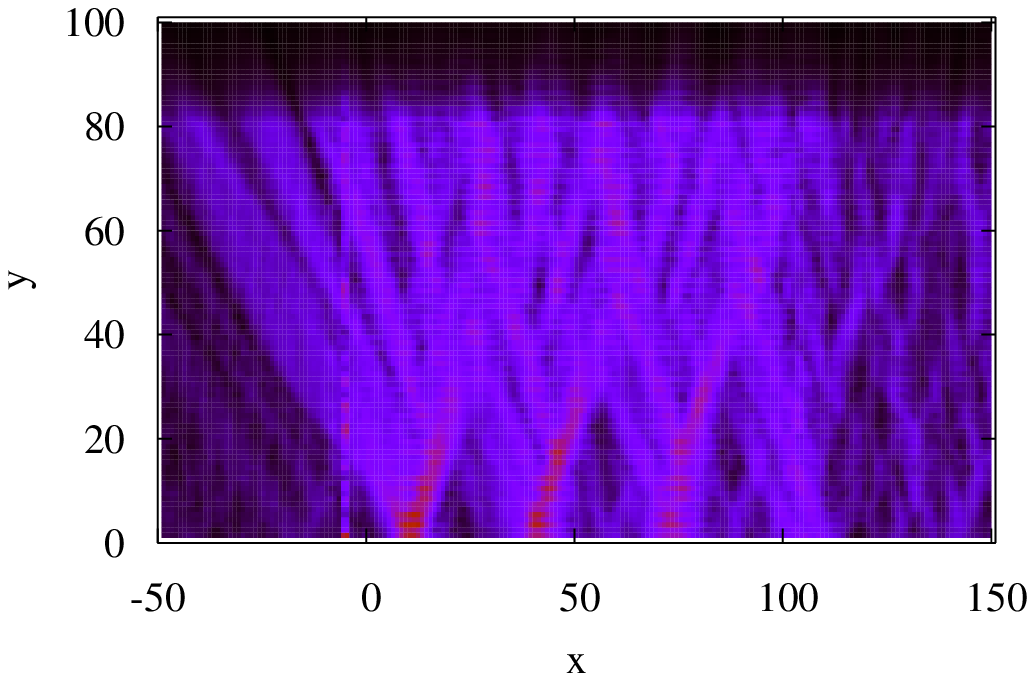}
\vspace{-2mm}\caption{\label{wfns} (color online) 
From top to bottom the electron (left) and hole (right) probability 
amplitudes are plotted, corresponding to 
the Andreev-transmission peaks marked by $A$ and $B$ 
in Fig.~\ref{trans1}, respectively. 
In our geometry $W_N = 80 $, $W = 100 $, lead $1$ (lead $2$) 
is located at $-10 < x < 0$ ($100 < x < 110$). 
Distances are in units of the lattice constant~\cite{parameters:dat}. 
}
\end{figure}

To reinforce the above detailed classical picture, 
we now calculate the electron and hole component of the 
wave functions inside the N/S
waveguide, and compare them with classical orbits. 
The contribution of the $n$-th incoming mode (from the left N-lead of width $W_L$) to the wave function 
at point ${\bf r}_S = (x,y)$ of the waveguide is 
\begin{equation}
\psi_n^{e(h)}({\bf r}_S) =  
\sum_{{\bf r}_L}  G^{ee(he)}({\bf r}_S,{\bf r}_L) \, \chi_n^{+,e}({\bf r}_L), 
\end{equation}
where ${\bf r}_L$ runs over the surface of the left lead. 
Here the appropriate components of the retarded Green's function are  
defined in Eq.~(\ref{BdG}), 
and $ \chi_n^{+,e}({\bf r}_L)$ is the transverse wavefunction
of the $n$-th incoming electron channel of the left lead, 
normalised to unit flux.   
The modulus square of the electron and hole components of the 
wave function  are shown in Fig.~\ref{wfns} for two different magnetic fields,
corresponding to the positive peaks $A$ and $B$ in Fig.~\ref{trans1}. 
For these scattering states, the hole probability amplitude 
has a local maximum at lead $2$. 
There are several other maxima of the probability amplitudes of the 
wave function at the lower side of the waveguide, both for the holes and for the electrons.
For each positive peak of ($T_a - T_0$), the condition Eq.~(\ref{cond:eq})  is satisfied,
where  $n$ is the number of maxima of the hole probability amplitude between the leads, 
and $\delta$ is the distance between the nearest electron and hole maxima. 

In conclusion, we have shown that even in the absence of ferromagnetic leads, 
an enhanced non-local current can be obtained by including a normal 
region between the leads and superconductor, and applying magnetic
fields perpendicular to the system. 
The current flowing from lead $1$ to lead $2$ shows 
oscillations with alternating signs as a function of magnetic field 
in the small-field regime, corresponding to alternating magnetic 
focusing of electron and hole-like quasi-particles between the two leads.  
Unlike an earlier proposal~\cite{feinberg}, where $T_a$ is exponentially 
suppressed with lead separation, 
the non-local current remains significant even for a lead
separation much bigger than the coherence length of the
superconductor. 
We discussed how the quantum results could be interpreted
qualitatively in a fully classical treatment providing a better
insight into the Andreev drag effect in our system. 
For the future it would be of interest to extend the
semi-classical approach developed for normal focusing problem. 
In this analysis one has to involve the semi-classical 
wave functions of the particles taking into account the more complex  
caustics~\cite{vanhouten} formed for both electrons and holes. 

We would like to thank A. F. Morpurgo, C. W. J. Beenakker 
and A. Korm\'anyos for useful discussions.
This work is supported by E. C. Contract No. MRTN-CT-2003-504574, EPSRC,  
the Hungarian-British TeT, 
and the Hungarian  Science Foundation OTKA TO34832.


\begin{thebibliography}{99}
\bibitem{parameters:dat}
Throughout the paper 
we set $\gamma=1$, $\epsilon_0=1.5$, $W_N=40a$ and $\Delta_0= 0.4$. 
For having one open channel in the leads $1$ and $2$ 
we set $\epsilon_0=3.8$ in the leads of width $W_L=10a$.
The Fermi energy is $E_F = 4 \gamma - \epsilon_0$. The separation between leads is $L=100a$.
The lattice constant 
is $a=\left(\lambda_F/2\pi\right) \sqrt{E_F/\gamma}$.
\end{thebibliography}
\end{document}